\documentclass{article}
\usepackage[T1]{fontenc} % add special characters (e.g., umlaute)
\usepackage[utf8]{inputenc} % set utf-8 as default input encoding
\usepackage{ismir,amsmath,cite,url}
\usepackage{graphicx}
\usepackage{color}
\usepackage{amssymb}
\usepackage{booktabs}
\usepackage[hidelinks,colorlinks=true,bookmarks=false]{hyperref}
\usepackage[table]{xcolor}
\usepackage{colortbl}
\usepackage{caption}
\usepackage{amsmath}
\usepackage{graphicx}
\usepackage{lineno}
\usepackage{microtype}
\usepackage{enumitem}
\usepackage{footmisc}
\usepackage{inconsolata}
\definecolor{myblue}{rgb}{0,0.0,0.0}
\hypersetup{linkcolor=myblue,urlcolor=myblue,citecolor=myblue,anchorcolor=myblue}

\title{Nested Music Transformer: Sequentially Decoding Compound Tokens in Symbolic Music and Audio Generation}

\multauthor
  {HaeJun Yoo$^1$ \hspace{1cm} Hao-Wen Dong$^2$ \hspace{1cm}  Jongmin Jung$^1$ \hspace{1cm}  Dasaem Jeong$^3$} {
  $^1$Dept. of Artificial Intelligence, $^3$Dept. of Art \& Technology, Sogang University, Seoul, South Korea\\
  $^2$University of California San Diego, US\\
  {\tt\small \{judejiwoo, jongmin, dasaemj\}@sogang.ac.kr, hwdong@ucsd.edu}}

\def\authorname{J. Ryu, H-W. Dong, J. Jung, and D. Jeong}

\begin{document}

\maketitle
\begin{abstract}
Representing symbolic music with compound tokens, where each token consists of several different sub-tokens representing a distinct musical feature or attribute, offers the advantage of reducing sequence length. While previous research has validated the efficacy of compound tokens in music sequence modeling, predicting all sub-tokens simultaneously can lead to suboptimal results as it may not fully capture the interdependencies between them. We introduce the Nested Music Transformer (NMT), an architecture tailored for decoding compound tokens autoregressively, similar to processing flattened tokens, but with low memory usage. The NMT consists of two transformers: the main decoder that models a sequence of compound tokens and the sub-decoder for modeling sub-tokens of each compound token. 
The experiment results showed that applying the NMT to compound tokens can enhance the performance in terms of better perplexity in processing various symbolic music datasets and discrete audio tokens from the MAESTRO dataset.

\end{abstract}

\begin{figure}{
 \includegraphics[width=\linewidth]{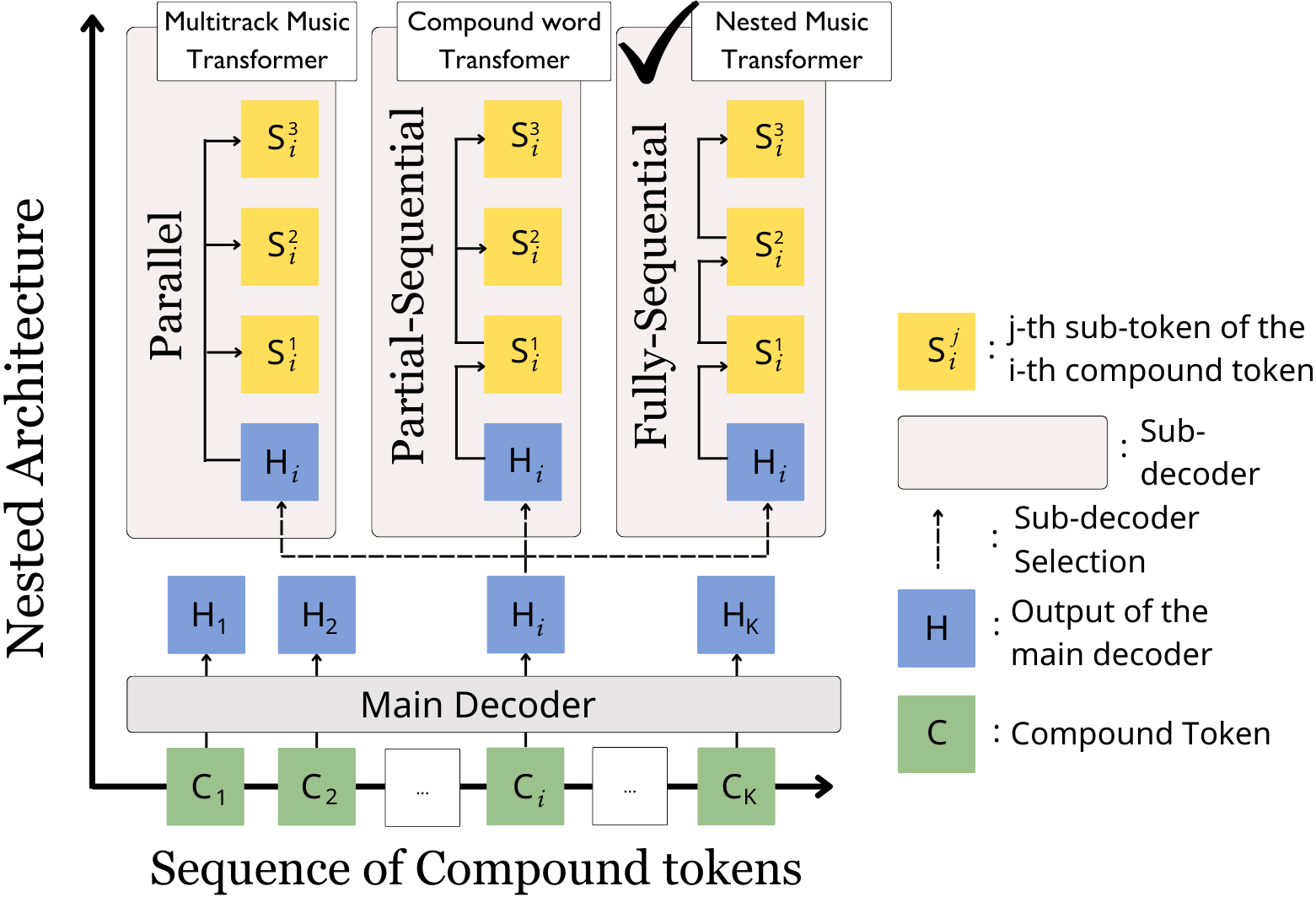}}
 \caption{Diagram of the nested architecture with three different methods for predicting sub-tokens.} %Our proposed NMT architecture predicts each sub-token in a fully sequential manner.}
 \label{fig:nmt_teaser}
\end{figure}

\begin{figure*}{
 \includegraphics[width=\linewidth]{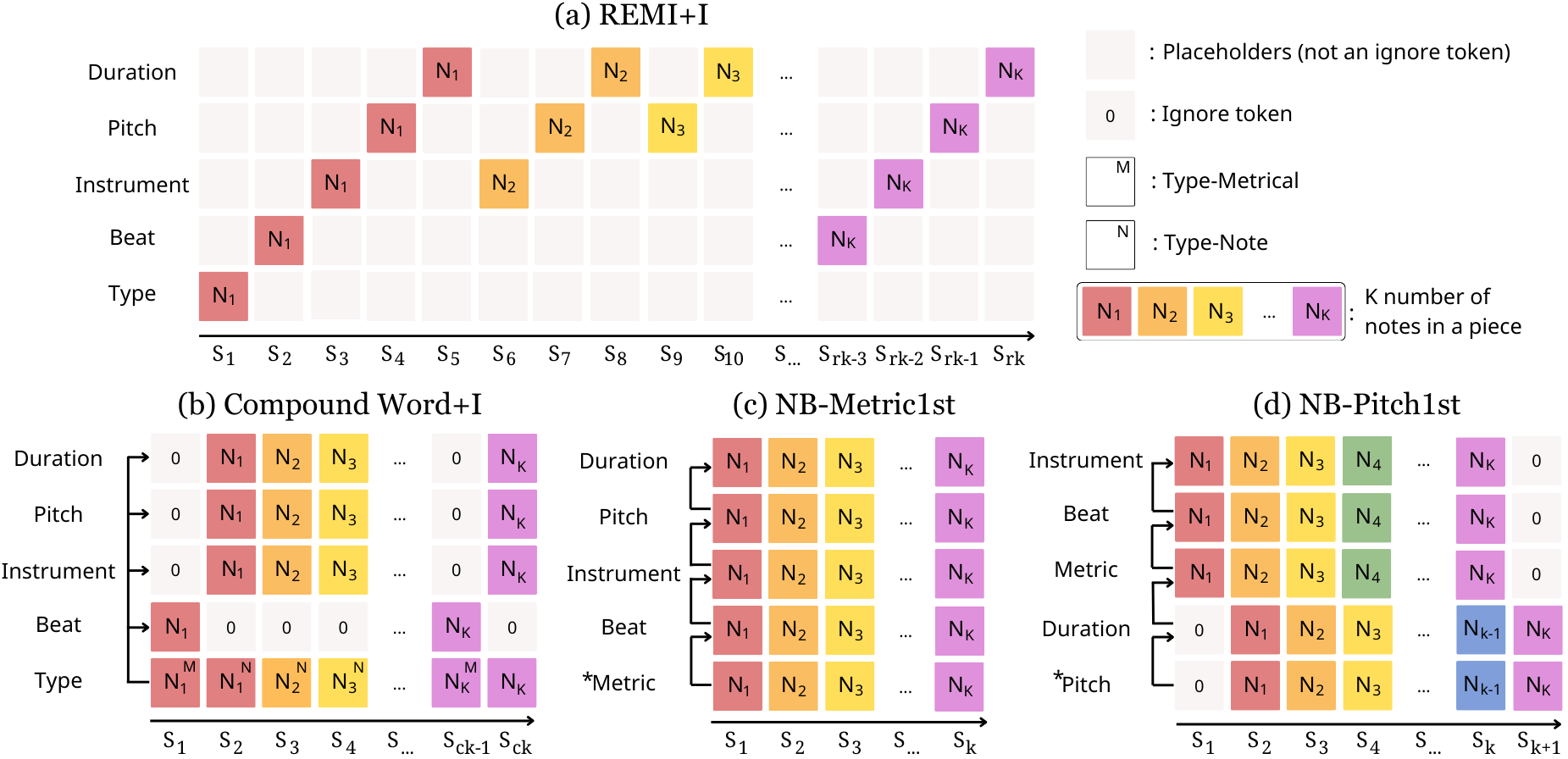}}
 \caption{An example illustrating the proposed representations, note-based (NB) encoding (c) NB-Metric1st and (d) NB-Pitch1st, alongside REMI and Compound word. All encodings represent the same piece of music by using five musical features. Specifically, REMI and Compound word were not originally designed for multi-instrument pieces, which is why we renamed the encodings with ``+I'' to (a) and (b). Here, $k$ denotes the number of notes and sequence length for NB, while $r$ and $c$ represent the ratios for REMI and Compound word, with values greater than 1.}
 \label{fig:encoding scheme}
\end{figure*}

\section{Introduction}\label{sec:introduction}

The effectiveness of the autoregressive language model becomes dominant in generative tasks in various domains, including music. The language model has been the most widely used generative model in symbolic music generation~\cite{huang2019musictransformer, ren2020popmag, huang2020pop, hsiao2021compound}. After the success of vector quantization or residual vector quantization~\cite{defossez2022high}, the language model is also widely applied to audio-domain music generation~\cite{NEURIPS2023_94b472a1,le2024stack, yanguniaudio}.

The power of the language model comes from its autoregressive modeling of sequential information. Once the data is \textit{flattened} to a sequence of discrete tokens, the language model can be applied in a straightforward manner. There have been many successive works on representing symbolic music data in a sequence of flattened tokens, such as MIDI-like encoding \cite{hawthorne2018enabling} or REMI \cite{huang2020pop}. 

However, a limitation of this approach is that the sequence length is quite lengthy, with the average number of tokens for pieces within the Lakh MIDI dataset~\cite{raffel2016learning} reaching 14,647. To overcome this limitation, the Compound Word Transformer \cite{hsiao2021compound} proposed an encoding scheme named Compound word that represents symbolic music as a sequence of compound tokens, in which several musical features or attributes are encoded into a single multi-dimensional token. 
%By grouping metric-related musical features such as position, tempo, and chord into a ``type-metrical'' group, and grouping note-related features like pitch, duration, and velocity into a ``type-note'' group, 
By grouping musical features into two different compound token types, metric and note, Compound word shortens the sequence length to less than half of what is encoded with REMI
%as depicted in Table~\ref{table:symbolic_nll_comparison}.
as depicted in Figure~\ref{fig:encoding scheme}. Similarly, Multitrack Music Transformer \cite{dong2023multitrack} employed a compound token scheme that encodes 
beat position, instrument, pitch, and duration into a single token,
% all possible musical features into a single group, 
resulting in a sequence length approximately one-third of that encoded with REMI. 
Furthermore, note-level compound tokens demonstrated a clear advantage in performance for discriminative tasks such as identifying the genre or style of music and suggesting accompaniments~\cite{zeng2021musicbert}.

Despite these attempts to reduce the sequence length by packing musical features into a single compound token for various purposes, encoding schemes which flatten tokens like REMI are still dominant in symbolic music generation. Both \cite{hsiao2021compound} and \cite{dong2023multitrack} in symbolic music generation showed that the generation with REMI was favored in their listening tests. One of the causes is that the previous models are designed to predict multiple features %from a single compound token 
in a parallel~\cite{dong2023multitrack} or partial-sequential~\cite{hsiao2021compound} way without considering interdependencies between different musical features encoded within the compound token, as depicted in Figure~\ref{fig:nmt_teaser}. 

%It is noteworthy that in discriminative tasks such as genre or style classification or accompaniment suggestion, note-level compound tokens like OctupleMIDI show a clear advantage in performance gain \cite{zeng2021musicbert}, as they do not have to generate tokens autoregressively.
%This implies that the main limitation of compound music token comes from 

To address this challenge, we introduce a novel decoding framework called the Nested Music Transformer (NMT). %illustrated in Figure~\ref{fig:nmt_teaser}. 
The primary goal of this framework is to decode compound tokens in a fully sequential manner while maintaining efficient memory usage. The proposed NMT combines two distinct cross-attention architectures within its sub-decoder: the intra-token decoder and the Embedding Enricher. The intra-token decoder autoregressively decodes the sub-tokens of a single compound token, while the Embedding Enricher updates embedding of each sub-token by attending to the hidden states of previous compound tokens. 
% Embedding Enricher is designed to refine key/value pairs (embeddings of sub-tokens) with contextual information before they are attended to by the query (output of the main decoder) as depicted in Figure~\ref{fig:sub-decoders}. 
% This refinement helps to balance the depth of the key/value and query vectors, thereby enhancing the efficiency of the cross-attention process.
%This enables the cross-attention to be implemented effectively by utilizing the context-enriched key/value pairs.
%combines a cross-attention architecture with an enrich-attention mechanism, enhancing the sequence of features (key/value) with the contextual information. This enables the cross-attention architecture to effectively utilize enriched key/value pairs.

We demonstrated that our proposed architecture achieves performance comparable to that of flattening-based models, while requiring fewer computational resources in terms of GPU memory and training time. This was confirmed through both quantitative evaluations and subjective listening tests for symbolic music generation. Furthermore, our experiments showed that the NMT and other nested architectures perform similarly to strong baseline models when generating audio samples using discrete audio tokens. All source code, pretrained models and generated samples are available at \url{https://github.com/JudeJiwoo/nmt}.

\section{Note-based Encoding}
Before we introduce the Nested Music Transformer, we explain Note-based encoding (NB), a compound token encoding scheme that we utilized as the primary encoding method. NB stands out for its ability to encapsulate the most comprehensive set of musical features within a single compound token, as illustrated in Figure~\ref{fig:encoding scheme}.

\subsection{Musical Features in Symbolic Encoding}
As depicted in Figure~\ref{fig:encoding scheme}, REMI, Compound word, and NB utilize several musical features to represent music pieces. We used a total of eight features: beat (position), pitch, and duration were essential, while instrument, chord, tempo, and velocity were selectively included based on the dataset characteristics. To encode other information, such as measure boundary and change in time signature, we also employed one additional feature \textit{Type} or \textit{Metric} following ~\cite{hsiao2021compound}.

%The \textit{Type} feature in compound token has several more roles than that of REMI. In Compound word (CP), there are two extra special tokens called ``metrical'' and ``note''. These tokens show whether the compound token only contains metric information like beat, chord, and tempo or note information such as pitch, duration, and velocity.
% includes information about beat, chord, and tempo (for "metrical") or about instrument, pitch, duration, and velocity (for "note").
In Compound word (CP), musical features are categorized into two groups: ``metrical'' and ``note.'' Consequently, the encoding employs two \textit{Type} tokens to specify the group of each compound token.
%In contrast to CP, NB always represents one single note as a single compound token, similar to Octuple MIDI~\cite{zeng2021musicbert}. This approach assigns a beat position to every note token, which leads to frequent repetition of beat information.
% In contrast to Compound word, NB combines all possible features into a single compound token, which results in the repetition of beat features too, marked as ``Conti token'' in Figure~\ref{fig:encoding scheme}. 
% To address the inefficiency caused by repeated beat information, we utilize the \textit{type} sub-token to encode changes in metrical structure.
Unlike CP, NB does not require group indicator tokens however, since each note token in NB is assigned a beat, unlike REMI and CP, we designed the \textit{Metric} feature to encode changes in the metrical structure. This allows the model to efficiently represent metrical changes within a single sub-token.
Specifically, the \textit{Metric} feature indicates whether the current note introduces a new time signature, measure, or beat, or continues the previous metrical context. For this purpose, we define four distinct values for the \textit{Metric} feature vocabulary, each representing a different combination of metrical changes or continuations.

The \textit{Beat} indicates the relative position of each note within a measure. %We carefully selected time signatures to correctly represent position of notes.
%To manage this, we set a restriction on the time signature, limiting it to no more than two whole notes. For time signatures exceeding this limit or those in unconventional formats, such as 33/8, we divided them according to a standard time signature we established.
The \textit{Chord} was derived using a rule-based algorithm from~\cite{hsiao2021compound}. %For chord analysis, we only considered notes with pitches in the range of 21 to 108, excluding any percussion-related instruments.
The \textit{Tempo} was set to follow an exponential scale for value changes, with this application varying across datasets.
%with a maximum value and different set of bins are created according to whether the MIDI is expressive or score-based.
The \textit{Instrument} feature specifies the instrument playing the note. In order to keep the variety of instruments manageable, we adopted the approach suggested in~\cite{dong2023multitrack}, trimming to 61
% different
types of instruments. %limiting the vocabulary to a selection of 61 different types of instruments.
The \textit{Pitch} feature utilized 128 categories of pitch values represented in MIDI. 
%However, the range of pitch values varies from dataset to dataset (e.g., 86 pitch ranges exist in the Pop1k7 dataset), meaning each dataset has a different size of pitch vocabulary.
The \textit{Duration} refers to the length of time each note is played. %We have imposed a cap on note lengths. %also carefully selecting the possible bins of durations.
% restricting them to a maximum of 8 beats. Only a select range of typical note durations is permitted, and any durations that fall outside this predefined range are substituted with the nearest equivalent duration within our accepted set.
The \textit{Velocity} represents MIDI velocity (dynamics) of each note. 
%In expressive MIDI, 17 velocity bins were used as a predefined set.

For the NB encoding method, a music piece P with $K$ number of notes, $P = \{n_{1}, n_{2}, n_{3}, ..., n_{K}\}$ can be conceptualized as a sequence of compound tokens, denoted by $P_{nb} = \{x_{1}, x_{2}, x_{3}, \ldots, x_{K}\}$, wherein each event $x_{i}$ is a compound token comprising up to eight sub-tokens in the orders like followings: 
\begin{align*}
(x_i^{\text{metric}}, x_i^{\text{beat}}, x_i^{\text{chord}}, x_i^{\text{tempo}}, x_i^{\text{inst}}, x_i^{\text{pitch}}, x_i^{\text{dur}}, x_i^{\text{vel}})
\end{align*}

\subsection{Compound Shift}
%CP \cite{hsiao2021compound} aimed to optimize the performance of the transformer decoder by segregating tokens into those related to ``metrical'' aspects and those related to ``note'' attributes. Similarly, 
% By shifting how we group the sub-tokens into a single compound token, we can make the main decoder focus more on important sub-tokens like pitch. 
By reordering the sub-tokens within a compound token, we can position the target sub-token to be predicted first. This adjustment enhances the objective metric of the target sub-token, as it benefits from being processed primarily by the more powerful main decoder rather than the sub-decoder. Each event $x_{i}$ which is shifted to pitch-first option comprises features like following:
\begin{align*}
(x_{i-1}^{\text{pitch}}, x_{i-1}^{\text{dur}}, x_{i-1}^{\text{vel}}, x_i^{\text{metric}}, x_i^{\text{beat}}, x_i^{\text{chord}}, x_i^{\text{tempo}}, x_i^{\text{inst}})
\end{align*}
Note that the order of prediction of each sub-token in the entire \textit{flattened} sequence does not change, and only the grouping boundary for a single compound token is shifted as depicted in Figure~\ref{fig:encoding scheme} (d). %This shifting strategy enables the transformer to predict note-related features while its metric information is already decided, as depicted in Figure~\ref{fig:encoding scheme}. 
% this is mixture of CP and MMT
% because now we know in which beat we are predicting note tokens.
We will refer to the non-shifted representation as \textit{NB-MF} and the shifted version as \textit{NB-PF}.

\begin{figure*}{
 \includegraphics[width=\linewidth]{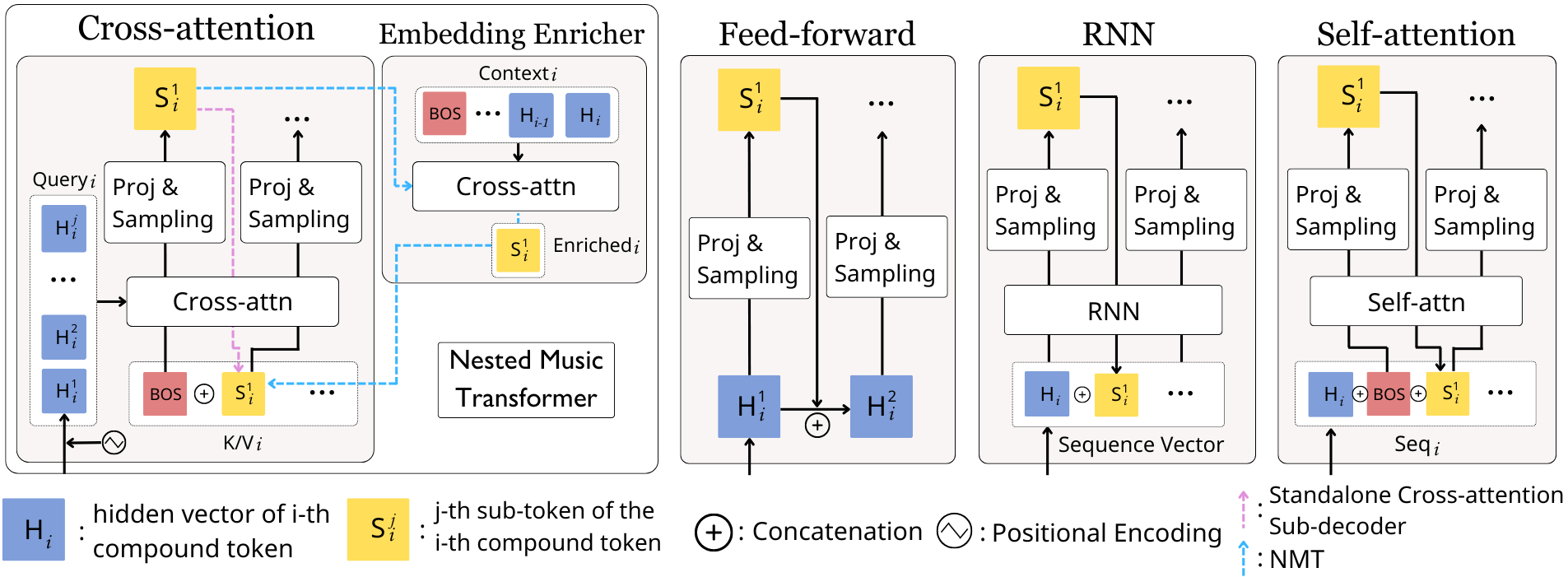}}
 \caption{Illustrations of the proposed Nested Music Transformer (NMT) and other sub-decoder structures}
 \label{fig:sub-decoders}
\end{figure*}

% describe overall architecture. eg: the model consists of three parts: token embedding, main decoder, and sub-decoder.
\section{Nested Music Transformer}
In this section, we introduce the architecture of Nested Music Transformer (NMT), which is designed to handle compound tokens. 
%In overall, the NMT follows the typical transformer-based language model which is decoder-only transformer\cite{liu2018generating, brown2020language}. 
%The structure is composed of three primary components: token embedding, main decoder, and sub-decoder. The token embedding summarizes the embeddings of each feature into a single vector. Subsequently, the main decoder processes the sequence of these vectors using a decoder-only transformer architecture. Lastly, the sub-decoder decodes out musical features from a given hidden vector which is the output of main decoder. The NMT employs an uniquely designed sub-decoder that incorporates cross-attention with a supportive structure called the Feature enricher as illustrated in Figure~\ref{fig:sub-decoders}.
The structure is composed of three primary components: token embedding, main decoder, and sub-decoder. The token embedding component summarizes the embeddings of each sub-token into a single vector which represents each compound token. Subsequently, the main decoder processes the sequence of these vectors using a decoder-only transformer architecture. Lastly, the sub-decoder decodes sub-tokens from the output of the main decoder. 
The proposed NMT integrates two distinct cross-attention architectures within its sub-decoder: the intra-token decoder and the Embedding Enricher. As the NMT generates sub-tokens, their embeddings are updated with contextual information by the Embedding Enricher, as illustrated in Figure~\ref{fig:sub-decoders}.
%The NMT employs a uniquely designed sub-decoder that incorporates cross-attention based sub-token decoder with an cross-attention architecture for embedding enhancement purpose called the Embedding Enricher, as illustrated in Figure~\ref{fig:sub-decoders}.

\subsection{Token Embedding \& Main Decoder}
%The \textit{token embedding} component involves multiple embedding layers for each feature, which deviates from the traditional approach of employing a single embedding layer for the entire vocabulary. 
%The token embedding module is designed to generate an embedding for each compound token in the input sequence. 
To summarize multiple embeddings from each sub-token, we simply sum them along the sub-token axis following \cite{dong2023multitrack, NEURIPS2023_94b472a1}. Additionally, we integrate learnable absolute positional embedding \cite{gehring2017convolutional} to denote the position of compound tokens within the sequence.
%The \textit{main-decoder} consists of multiple causal self-attention layers, denoted as Self-attention(*) below. 
% For a specified set of compound tokens $P_{nb} = \{x_{1}, x_{2}, x_{3}, \ldots, x_{K}\}$, 
Specifically, the $i$-th compound token $x_i$ in the sequence is converted into a vector through the token embedding process and aggregated with its positional embedding. This combined vector is then fed into the main decoder, producing the output of the main decoder, also known as the hidden vector $h_i$.

\subsection{Sub-decoder with Cross Attention}
%The sub-decoder is in take charge of decoding this hidden vector $h_i$ into musical features by sequentially creating logits and probability distributions. 
The main goal of the sub-decoder is to obtain proper hidden state to predict output sub-token $s_i^{j}$ 
which is $j$-th sub-token of $i$-th compound token, based on output of the main decoder $h_i$ and the preceding output sub-tokens $s_i^0, \dots, s_i^{j-1}$ that are predicted before.

Many previous works have suggested using a similar sub-decoder to sequentially predict the sub-token sequence, such as updating hidden state by concatenating with the embedding of sub-tokens \cite{hsiao2021compound}, using RNN \cite{wang2020pianotree} or causal self-attention \cite{yanguniaudio}. However, through comparative experiments presented in Section~\ref{sec:experiments}, we found that applying cross-attention is one of the most effective way to model the compound token sequence in symbolic music. 

% \subsubsection{Cross-attention Strategy}
The cross-attention-based sub-decoder operates by iteratively concatenating a key/value pair sequence $\text{K/V}_i$ with embeddings of sub-tokens $\text{Emb}(s_i)$, %along the sub-token sequence axis, 
starting with an initial key/value sequence that contains only the beginning-of-sequence $BOS$ token. For each sub-token to be sampled, the architecture computes multi-head scaled dot-product cross-attention between the query sequence, consisting of %duplicated and 
positionally encoded output of the main decoder $h_i$, and the current key/value sequence. The positional encoding of $h_i$ ensures that the hidden vector has a distinct bias for predicting target sub-token. From the attention output $a_i^{j}$, the matrix $\text{W}_\text{logits}^{j}$ is applied to create logits. This iterative process continues until all sub-tokens are sampled. The process can be expressed as follows:
% \begin{align}
%     \text{Query}_i &= \text{Duplicate}(h_i, C), \\
%     \text{Query}_i &= \text{PositionalEncoding}(\text{Query}_i), \\
%     \text{K/V}_i^{f_1} &= BOS, \\
%    a_{\text{i}}^{f_c} &= \text{Cross-Attention}(\text{Query}_i, \text{K/V}_i^{f_{c-1}}), \\
%     \hat{o}_i^{f_c} &= \text{Sampling}(\text{Softmax}(a_{\text{i}}^{f_c}W_\text{logits}^{f_c})), \\
%     \text{K/V}_i^{f_c} &= \text{K/V}_i^{f_{c-1}} \oplus \text{Embedding}(\hat{o}_i^{f_c})
% \end{align}
\begin{align}
    \text{Query}_i^j &= \text{PositionalEncoding}(h_i), \\
    %\text{Query}_i &= \text{PositionalEncoding}(\text{Query}_i), \\
    % \text{K/V}_i^{0} &= BOS, \\
    % \text{K/V}_i^{c} &= \text{Concat}(BOS, \text{Emb}(o_i^0), ..., \text{Emb}(o_i^{c-1})), \\
    \text{K/V}_i^{j} &=
    \begin{cases}
        BOS & \text{if } j = 0, \\
        \text{Concat}(BOS, \ldots, \text{Emb}(s_i^{j-1})) & \text{if } j > 0, \\
    \end{cases}, \\
    a_i^{j} &= \text{Cross-Attention}(\text{Query}_i^j, \text{K/V}_i^{j}), \\
    s_i^j &= \text{Sampling}(\text{Softmax}(a_i^{j}\text{W}_\text{logits}^j))
\end{align}
%Here, the Duplicate function is used to replicate the output of the main decoder, $h_i$, into the number of sub-tokens. Each duplicated $h_i$ is then positionally encoded, ensuring that each hidden vector has a different bias for predicting different sub-tokens. 

% The idea of applying cross-attention instead of self-attention 

\subsubsection{Embedding Enricher}
Since the embedding of a sub-token is a shallower vector compared to the output of the main decoder, we designed a cross-attention architecture called the Embedding Enricher. This architecture updates embedding of sub-token $\text{Emb}(s_i)$ with a context sequence derived from the prior outputs of the main decoder $h_{i-(w-1)}, ..., h_i$, where $w$ represents the window size.
%This architecture is an ad-hoc enhancement to CD, enriching the sub-token sequence in CD with context information from prior hidden states or outputs of the main decoder. In this attention process, the key/value sequence in CD becomes the query $\text{Query}_i$, and the prior hidden states $h_{i-(w-1)}, ..., h_i$, where $w$ represents the window size, serve as the key/value sequence $\text{K/V}_i$. The process can be represented as follows:
%The feature enricher architecture enhances the embedding vectors of individual features used as key/value sequences for cross-attention. This architecture progressively refines the feature sequence via enrich-attention, a process that directs attention towards prior hidden states, thereby effectively incorporating context information into the feature sequence. Following this process, the augmented key/value sequences are employed within the cross-attention framework to derive attention-weighted representations. The process can be represented as follows:
% \setcounter{equation}{0}
% \begin{align}
% \text{Features}_i^{f_1} &= \text{BOS}_\text{feature}, \\
% \text{K/V}_i &= \text{BOS}_\text{k/v} \oplus h_{i-(w-1)} \oplus h_{i-(w-2)}, ..., \oplus h_i, \\
% \text{K/V+}_i^{f_c} &= \text{Enrich-Attention}(\text{Features}^{f_{c-1}}, \text{K/V}_i^{f}), \\
% a_{\text{i}}^{f_c} &= \text{Cross-Attention}(\text{Query}_i, \text{K/V+}_i^{f_c}), \\
% \hat{o}_i^{f_c} &= \text{Sampling}(\text{Softmax}(a_i^{f_c}\textbf{W}_\text{logits}^{f_c})), \\
% \text{Features}_i^{f_c} &= \text{Features}_i^{f_{c-1}} \oplus \text{Embedding}(\hat{o}_i^{f_c})
% \end{align}
\begin{align}
    % \text{Query}_i^{c-1} &= \text{Concat}(BOS_{\text{query}}, \text{Emb}(o_i^1), ..., \text{Emb}(o_i^{c-1})), \\
    % \text{Query}_i^{c} &=
    % \begin{cases}
    %     BOS & \text{if } c = 0, \\
    %     \text{Concat}(BOS, \ldots, \text{Emb}(o_i^{c-1})) & \text{if } c > 0, \\
    % \end{cases}, \\
    % \text{E}_i &= \text{Emb}(o_i), \\
    \text{Context}_i &= \text{Concat}(BOS, h_{i-(w-1)}, ..., h_i), \\
    \text{Enriched}_i &= \text{Cross-Attention}(\text{Emb}(s_i), \text{Context}_i)
    % \text{K/V+}_i^{c} &= \text{cross-attention}(\text{Query}_i^{c}, \text{K/V}_i)
\end{align}
In the Nested Music Transformer, the output vector $\text{Enriched}_i$ replaces the original embedding of sub-tokens before being concatenated into the key/value pair sequence in Equation (2) as depicted in Figure~\ref{fig:sub-decoders}. These context-enriched embeddings allow the architecture to process attention with deeper vectors than the original embeddings, resulting in better performance on the objective metric compared to the standalone cross-attention-based sub-decoder, as demonstrated in Table~\ref{table:symbolic_nll_comparison}.
%where $w$ represents the window size for output of the main decoder or hidden vector $h$ which are utilized as context information for key/value pair sequence. 
%The output vector $\text{Enriched}_i$ replaces the original embedding of sub-tokens before concatenation into the key/value pair sequence in Equation (2) as depicted in Figure~\ref{fig:sub-decoders} in the Nested Music Transformer. These context-enriched embeddings enable the architecture to process attention with deeper vectors than original embeddings, resulting in better performance on the objective metric compared to the stand-alone cross-attention-based sub-decoder, as demonstrated in Table~\ref{table:symbolic_nll_comparison}.
%combines two cross-attention-based architectures designed to derive the hidden state for decoding sub-tokens from the context-enriched key and value sequence.
%This allows the cross-attention based architecture to process the attention with context-enriched key and value sequence. 
%The output of the attention K/V+ are subsequently employed as a new key/value sequence in CD, allowing the architecture to process the attention with enriched key and value vectors.

\subsection{Other Comparative Structures}
\subsubsection{Feed-forward-based Architecture}
% Catvec strategy is similar to the one presented in \cite{hsiao2021compound}. It updates the hidden state $h_i^{f_c}$ by concatenating previous hidden state and embedding of sampled output $o_i^{f_{c-1}}$.
The Feed-forward-based sub-decoder, inspired by \cite{hsiao2021compound}, iteratively updates the output of the main decoder to predict sub-tokens. It concatenates the previously used hidden state with the embedding of the last sampled output to predict the next sub-token. %followed by the projection layers which convert the concatenated vectors to match the main decoder's dimension.
\subsubsection{RNN-based Architecture}
RNN-based sub-decoder capitalizes on the sequential nature of recurrent neural network to update hidden state. The initial input sequence and hidden state utilize the output of the main decoder $h_i$, and through the iteration the embedding of the sampled output is appended to the input sequence until all the sub-tokens are generated.
\subsubsection{Self-attention-based Architecture\protect\footnote{The proposed self-attention-based sub-decoder operates differently from the method described in \cite{yanguniaudio}. % In our design, the output of the main decoder $h_i$ is treated as a single vector among the sequence. 
% for self-attention process
% , allowing the attention mechanism to utilize this information as needed for generating the next sub-tokens. 
%In contrast, 
Unlike ours, \cite{yanguniaudio} used $h_i$ as a base of every vector in the sequence, which is updated by the embedding of generated sub-tokens, similar to the operation of our proposed cross-attention-based sub-decoder. Experimental results indicate that the architecture in \cite{yanguniaudio} outperforms our self-attention-based architecture and delivers comparable results to our cross-attention-based architecture.}}
% The self-attention architecture (SA) aims to update the input sequence with sampled output $o_i$ from the previous hidden state or attention output $a_i^{c-1}$. To ensure that the initial attention values can be properly processed, we include a BOS token at the beginning of the sequence. The process can be summarized as follows:
The self-attention-based sub-decoder aims to get the sequence vector $\text{Seq}_{i}$ by iteratively concatenaing it with the embeddings of the sampled output $\text{Emb}(s_i)$. %from the attention output $a_i$. 
The initial sequence vector consists of the output of the main decoder $h_i$ and BOS token to ensure that the initial attention values can be properly processed. This sequence vector $\text{Seq}_{i}$ is then used as the query, key, and value in the self-attention mechanism. %which creates a new sequence axis by merging the previous batch $B$ and sequence $T$ into a new batch axis. To ensure that the initial attention values can be properly processed, we include a BOS token after reshaped hidden vector $h_i^{re}$. Since the attention output includes the hidden states of all sub-tokens, it is necessary to select the appropriate vector after self-attention is applied. 
The process can be summarized as follows:
% \begin{align}
% \text{Sequence}_{i}^{1} &= \text{Reshape}(h_i,(B \times T, 1, d)) \oplus BOS, \\
% a_\text{i}^{c-1} &= \text{Self-Attention}(\text{Sequence}_i^{c-1}), \\
% o_i^{c} &= \text{Sampling}(\text{Softmax}(a_\text{i}^{c-1}\text{W}_\text{logits}^{c-1})), \\
% \text{Sequence}_{i}^{c} &= \text{Sequence}_i^{c-1} \oplus \text{Emb}(o_i^{c})
% \end{align}
\begin{align}
%h_i^{`} &= \text{Reshape}(h_i,(B \times T, 1, d)), \\
% \text{Sequence}_{i}^{c} &= \text{Concat}(h_i^{re}, BOS, ..., \text{Emb}(o_i^{c-1}), \\
 \text{Seq}_{i}^{j} &=
\begin{cases}
    \text{Concat}(h_i, BOS) & \text{if } j = 0, \\
    \text{Concat}(h_i, BOS, ..., \text{Emb}(s_i^{j-1})) & \text{if } j > 0, \\
\end{cases}, \\
a_i^{j} &= \text{Self-Attention}(\text{Seq}_i^{j}), \\
s_i^{j} &= \text{Sampling}(\text{Softmax}(a_i^{j}\text{W}_\text{logits}^{j}))
% \text{Sequence}_{i}^{c} &= \text{Sequence}_i^{c-1} \oplus \text{Emb}(o_i^{j})
\end{align}
%Here, $h_i$ and $BOS$ are concatenated to form the input sequence, and $a$ represents the output of attention from the self-attention mechanism.

%The self-attention architecture operates by initially preparing an input sequence consisting of the initial hidden vector $h_i$ and a special token denoting the beginning of the feature sequence $BOS$. This sequence undergoes self-attention, wherein each element attends to all others to compute a weighted representation capturing the relevance of each element for predicting the next feature. The resulting attention vector highlights important information for feature prediction. The embedding of sampled feature acquired from the vector is then appended to the input sequence, creating an updated sequence. The process of self-attention can be represented as follows:

\begin{table*}[htbp]
\centering
\scriptsize
\setlength\tabcolsep{3.7pt} % Adjusting column separation
\begin{tabular}{lccccccccccccccc}
\toprule
 & \multicolumn{6}{c}{SOD} & \multicolumn{3}{c}{Lakh} & \multicolumn{3}{c}{Pop1k7} & \multicolumn{3}{c}{Pop909} \\
\cmidrule(lr){2-7} \cmidrule(lr){8-10} \cmidrule(lr){11-13} \cmidrule(lr){14-16}
 & GPU mem.(GB) & Time(s) / iter. & Token Len. & Mean$\downarrow$ & Beat & Pitch & Mean & Beat & Pitch & Mean & Beat & Pitch & Mean & Beat & Pitch \\
\midrule
\rowcolor{white} REMI\cite{huang2020pop} & 19.90 & 0.461 & 6,638($\pm$7,518) & \textbf{0.474} & \textbf{0.229} & \textbf{0.753} & \textbf{0.294} & 0.293 & \textbf{0.408} & \textbf{1.087} & \textbf{0.470} & \textbf{1.138} & \textbf{0.716} & 0.368 & \underline{0.984} \\
\arrayrulecolor{black!40}\cmidrule(lr){1-16}
\rowcolor{white} CP\cite{hsiao2021compound} & 7.93 & 0.119 & 3,230($\pm$3,480) & 0.604 & 0.257 & 0.971 & 0.361 & 0.288 & 0.527 & 1.172 & 0.495 & 1.219 & 0.911 & 0.410 & 1.220 \\
CP* + NMT & 16.13 & 0.224 & -- & \underline{0.545} & \underline{0.237} & 0.864 & 0.327 & 0.288 & 0.466 & \underline{1.103} & \underline{0.483} & \underline{1.154} & \underline{0.724} & \underline{0.334} & \textbf{0.969} \\
\arrayrulecolor{black!40}\cmidrule(lr){1-16}
NB-MF + Par.\cite{dong2023multitrack} & 8.40 & 0.123 & 2,398($\pm$2,764) & 0.712 & 0.466 & 1.084 & 0.431 & 0.431 & 0.604 & 1.480 & 0.871 & 1.802 & 1.003 & 0.674 & 1.393 \\
NB-MF + NMT & 16.14 & 0.215 & -- & 0.567 & 0.246 & 0.906 & 0.324 & \textbf{0.276} & 0.466 & 1.168 & 0.503 & 1.304 & 0.803 & \textbf{0.264} & 1.114 \\
\arrayrulecolor{black!40}\cmidrule(lr){1-16}
NB-PF + Par. & 8.30 & 0.120 & -- & 0.632 & 0.565 & 0.913 & 0.376 & 0.502 & 0.481 & 1.396 & 0.998 & 1.604 & 0.986 & 0.824 & 1.359 \\
NB-PF + CA & 14.74 & 0.174 & -- & 0.564 & 0.276 & 0.867 & \underline{0.305} & 0.287 & \underline{0.424} & 1.161 & 0.538 & 1.244 & 0.767 & 0.357 & 1.052 \\
NB-PF + NMT & 16.13 & 0.217 & -- & 0.549 & 0.263 & \underline{0.855} & 0.306 & \underline{0.285} & 0.427 & 1.149 & 0.515 & 1.243 & 0.771 & 0.345 & 1.090 \\
\arrayrulecolor{black}\cmidrule(lr){1-16}
\rowcolor{white} NB-PF + FF & 8.12 & 0.122 & -- & 0.607 & 0.361 & 0.881 & 0.338 & 0.372 & 0.449 & 1.280 & 0.635 & 1.396 & 0.850 & 0.431 & 1.121 \\
\rowcolor{white} NB-PF + RNN & 9.77 & 0.144 & -- & 0.591 & 0.300 & 0.915 & 0.315 & 0.297 & 0.437 & 1.166 & 0.531 & 1.257 & 0.792 & 0.366 & 1.077 \\
\rowcolor{white} NB-PF + SA & 15.67 & 0.181 & -- & 0.574 & 0.287 & 0.902 & 0.311 & 0.287 & 0.431 & 1.204 & 0.553 & 1.320 & 0.849 & 0.417 & 1.150 \\
\bottomrule
\end{tabular}
\\[\smallskipamount]
\textbf{CP*}: Compound word representation\quad
\textbf{NB-MF}: metric-first NB\quad
\textbf{NB-PF}: pitch-first NB\quad
\textbf{NMT}: cross-attention-based sub-decoder + Embedding Enricher\quad
\textbf{CA}: cross-attention-based sub-decoder\quad
\textbf{FF}: Feed-forward-based sub-decoder\quad
\textbf{SA}: self-attention-based sub-decoder\quad
\caption{Model comparison on their average NLL loss for symbolic music. The GPU memory usage and iteration times for each model in SOD is included. Additionally, we included the average token length and standard deviation across all pieces in SOD.}
% \\ NB-TF: Type-first NB, NB-PF: Pithc-first NB, CA: cross-attention, SA: self-attention, NMT: CA + feature enricher}
\label{table:symbolic_nll_comparison}
\end{table*}

\subsection{Self-attention versus Cross-attention}
The preference for cross-attention over self-attention arises from the observation that the output of the main decoder, $h_i$, already contains sufficient information to predict sub-tokens, as demonstrated in the parallel prediction method used in the Multitrack Music Transformer \cite{dong2023multitrack}. 
On the other hand, the embedding of the sampled output is comparatively shallow, lacking the previous context despite having the same dimension as $h_i$. Additionally, since both attention layers use a residual connection for the vectors used as keys, utilizing $h_i$ as the key facilitates a direct gradient flow. Therefore, updating $h_i$ as the key with cross-attention can be more advantageous than updating the embedding of the sampled sub-token with self-attention.

%\subsection{Sampling Method}
%Unfortunately, note-based encoding suffers from a significant issue of degeneration. To address this, the model was prevented from repetitively generating phrases by utilizing top-p sampling and temperature in sampling. 

%\subsection{Shifted Token Grouping}
%Additionally, as evident in Figure 1, given the characteristics of NB, it is possible to determine the first feature to predict from the hidden vector output of the main decoder. By comparing the traditional \textit{type-first} with \textit{pitch-first}, it was observed that the performance related to features associated with notes improved in the pitch 1st approach as depicted in the table X. Consequently, an ensemble method was introduced, harmonizing the two models to generate features related to metrical aspects in type 1st and features related to notes in pitch 1st, alternating between the two models for generating a single token.

\subsection{Applying to Audio Tokens}
%Although initially designed for sequences of compound tokens in the symbolic music domain, the architecture decoding features is similar to  discrete audio tokens obtained through multi-layer residual vector quantization. The approach in proposed from 
MusicGen~\cite{NEURIPS2023_94b472a1} has employed a four-level residual vector quantization technique for a single token, which bears similarity to using four musical features or sub-tokens for compound tokens in symbolic music. Given that the optimal architecture, particularly for decoding compound features in a fully-sequential manner, is still being explored for audio tokens\cite{yanguniaudio}, we employed the Nested Music Transformer on discrete audio tokens to assess the potential of our proposed architecture.
%it can handle any sequence of compound tokens. As an example, discrete audio tokens from residual vector quantization can be represented in the same format. We 

\section{Experiments}\label{sec:experiments}
\subsection{Dataset Preparation}
We selected four datasets to conduct our experiments on symbolic music generation: Pop1k7~\cite{hsiao2021compound}, Pop909~\cite{wang2020pop909}, the Symbolic Orchestral Database (SOD)~\cite{crestel2018database}, and the clean version of the Lakh MIDI Dataset (LMD clean)~\cite{raffel2016learning}, which is free of data leakage problems. 
During preprocessing, MIDI files without a time signature or with excessive or insufficient length were filtered out, and we specifically selected pieces featuring a minimum of four instruments for LMD clean. Note quantization varied across datasets: twelve resolutions per beat for SOD and four resolutions per beat for the others. %Given SOD's finer resolution, 
We also filtered out MIDI files with expressive tempo and timing.
We split the prepared data, reserving 10\% for validation and 10\% for testing. %For symbolic music, random segmentation of tunes into predefined lengths was applied per training epoch. 
Additionally, augmentation techniques for pitch and chord involved random semitone shifts $s \in \mathbb{Z}$ within a range of $s \sim U(-5, 6)$.

\subsection{EnCodec for MAESTRO}
For discrete audio tokens, we prepared MAESTRO dataset~\cite{hawthorne2018enabling}, which has 200 hours piano performance audio files. We fine-tuned the audio tokenizer proposed by \cite{NEURIPS2023_94b472a1} with MAESTRO audio files to create sequences of discrete audio tokens, each with 30 seconds of length. The sampling rate of the token is 50 Hz, which means 30 seconds of audio is represented with 1500 audio tokens, each with 4 different codebooks.

\subsection{Model and Hyperparameter Configuration}
The baseline models for symbolic music generation are defined as follows: flattening for REMI~\cite{huang2020pop}, partial-sequential Feed-forward-based sub-decoder for Compound word~\cite{hsiao2021compound}, and parallel prediction with NB-MF~\cite{dong2023multitrack}. Additionally, the \textit{delay} method proposed by \cite{NEURIPS2023_94b472a1} is explored as a baseline for generating audio tokens, which utilizes rearranged residual vectors or sub-tokens in a parallel manner. In exploring both symbolic music and audio token generation, we conducted experiments using the Nested Music Transformer (NMT) and various sub-decoder architectures to assess the effectiveness of our proposed model. To ensure a fair comparison among these models, we aimed for a comparable number of model parameters, approximately 40 million for symbolic music and 62 million for discrete audio tokens\footnote{Both models have 8 attention heads and a dimension size of 512, with a single layer for all sub-decoder architectures and an additional single layer for the Embedding Enricher when using the NMT. However, they have a total of 12 and 15 decoder layers, respectively.}. To enhance efficiency in processing long sequences within the transformer architecture, we integrated Flash attention\cite{dao2022flashattention}. 

Training the model entailed 100K steps for symbolic music and 200k for discrete audio tokens, utilizing the AdamW optimizer~\cite{loshchilov2017decoupled}, with a segment batch size of 8 and 16 for each task, where $\beta_1$ were set to 0.9, $\beta_2$ to 0.95, and a gradient clipping threshold was set to 1.0. We implemented a cosine learning rate schedule with a warm-up phase of 2000 or 4000 steps for each task. During this warm-up phase, the learning rate gradually increased before reaching its maximum value $1\times\mathrm{e}^{-4}$. %Instead of employing early stopping, dataset-specific dropout rates were applied to address overfitting concerns. 
To address overfitting concerns, we applied dataset-specific dropout rates instead of using early stopping. These dropout rates were chosen to ensure that the optimal validation loss remained stable until the end of training. %We conducted training using either a single GPU or multiple GPUs combined for distributed data parallel. Each GPU had a memory capacity of 24GB, and 
We utilized mixed precision techniques.

\subsection{Quantitative Evaluation on Symbolic Music}
We evaluated the symbolic music generation task using the average negative log-likelihood (NLL). However, directly comparing the loss values across models using different encoding schemes posed challenges. To address this, we first adjusted the input sequence length for each encoding scheme to ensure that the NLL is derived from a similar amount of context regardless of the encoding scheme. Furthermore, instead of calculating the average NLL as done during the training steps, we calculated it based on the set of probabilities of tokens processed with full context. To achieve this, we used a moving-window method with a window size equal to the input sequence length to create a set of overlapping input sequences.

Secondly, we adjusted the probabilities for each sub-token in a compound token to account for discrepancies between REMI and other encoding schemes like CP and NB. REMI omits redundant tokens such as repetitive positions (beat), as depicted in Figure~\ref{fig:encoding scheme}. Thus, when predicting a new note, a model based on REMI must decide whether to add the note at a new position by predicting a new beat token, or to add the note at the same position by predicting a pitch token. In contrast, CP and NB, due to the nature of their encoding schemes, split this prediction into two steps: first, they determine the beat position, and then they predict the pitch. This means they have more prior information when predicting the pitch token since changes in beat are already fixed and provided as a condition. %To address this discrepancy, 
To adjust the probability of sub-tokens in NB and CP, which differ due to the discrepancy, we accumulated the probability of each sub-token to the next token in NB or CP if that sub-token was omitted in its corresponding REMI encoding. For example, when predicting a pitch token at the same beat, $P(\text{pitch} \mid \text{context})$ in REMI can be compared to $P(\text{same\_beat} \mid \text{context}) \times P(\text{pitch} \mid \text{context}, \text{same\_beat})$ in NB or CP.

From the results, we observe several key tendencies. First, applying the Nested Music Transformer (NMT) enhances the overall performance across all types of compound token encodings, including previously suggested schemes like CP and NB-MF (similar to \cite{dong2023multitrack}). Second, the NMT demonstrates a clear advantage in using the cross-attention-based sub-decoder and the Embedding Enricher compared to other baseline architectures.
Finally, our pitch-first NB (NB-PF) encoding outperforms the metric-first NB (NB-MF) encoding in predicting pitch. This is because the model can predict the next pitch feature through the main decoder by leveraging the previously inferred note position information. Conversely, NB-MF showed lower loss in beat prediction. This difference arises from which sub-token relationships are calculated through the main decoder instead of the sub-decoder. Overall, the results indicate that pitch-first token grouping is an efficient strategy.

\begin{table}[]
\centering
\scriptsize
\setlength\tabcolsep{4pt}
\begin{tabular}{lcccc}
\toprule
                   & FAD-uncon$\downarrow$ & FAD-cond$\downarrow$ & KLD$\downarrow$   & mean NLL$\downarrow$   \\
\midrule
Parallel           & 0.166      & 0.206     & 0.075  & 4.669  \\
Flatten            & 0.140      & \textbf{0.176}     & 0.068  & 4.482  \\
Delay \cite{NEURIPS2023_94b472a1} & 0.168      & 0.188     & 0.066  & 4.564  \\
\arrayrulecolor{black!40}\cmidrule(lr){1-5}
Self-attention        & \textbf{0.131}      & 0.186     & 0.074  & 4.353  \\
Cross-attention        & 0.145      & 0.190     & \textbf{0.065}  & \textbf{4.314}  \\
NMT & 0.165      & 0.198     & 0.067  & 4.318  \\
\arrayrulecolor{black}\bottomrule
\end{tabular}
\caption{Model comparison for discrete audio tokens}
\label{table:encodec_comparison}
\end{table}

\subsection{Quantitative Evaluation on Discrete Audio Tokens}

We evaluated models with discrete audio tokens using following metrics: Fréchet Audio Distance (FAD), Kullback-Leibler Divergence (KL), and the mean NLL loss over sequences. A lower FAD score suggests that the generated audio is more plausible. To mitigate sample number bias for the test set, we employed adaptive FAD as proposed by \cite{gui2024adapting}, along with CLAP\cite{wu2023large} embeddings for each sample. FAD scores were computed based on 500 unconditionally generated samples and 345 samples generated given prompts. Following \cite{NEURIPS2023_94b472a1}, we computed the KL-divergence over the probabilities of the labels between the original and the generated audio samples. Table~\ref{table:encodec_comparison} shows the evaluation results.

We observe that using a cross-attention-based sub-decoder or the NMT achieves better NLL compared to a self-attention-based sub-decoder. However, the tendency differs from that seen in symbolic music. Adding the Embedding Enricher did not significantly improve performance in the audio domain. We hypothesize that this disparity arises from the distinct characteristics of tokens in both domains. In the symbolic domain, each musical feature requires context to form sufficient semantic information, whereas each token in the audio domain, with a 2048 vocabulary size codebook, contains more standalone information. This observation suggests potential avenues for future research, such as exploring effective methods to integrate the semantic information of symbolic music with discrete audio tokens.

\subsection{Subjective Listening Test}

For the subjective listening test, we used the Symbolic Orchestral Database (SOD)~\cite{crestel2018database} to generate MIDI samples given four-measure prompts. We carefully selected eight prompts from the test split and generated continuation results using four different models: two baseline models (REMI and CP) and two proposed models (CP + NMT and NB-PF + NMT). We applied different sampling methods to each model.\footnote{During the generation process, we used nucleus sampling (top-p sampling) with $p = 0.99$. Our proposed models were sensitive to the choice of the temperature parameter, where an improperly selected temperature would result in excessive repetition regardless of encoding schemes. Therefore, we searched for the optimal temperature value for each model within the range of [1.0, 1.3] on the validation set.}
We conducted the test with 29 participants, asking them to evaluate the generated outputs based on three criteria: \textit{Coherence} (the naturalness of transitions), \textit{Richness} (the variety of harmony and rhythm), and \textit{Consistency} (the lack of errors in composition), as well as an \textit{Overall} rating for the perceptual quality of the samples as a whole.

As summarized in Table~\ref{table:symbolic_listening_comparison}, our proposed models generated samples of comparable quality to REMI, outperforming the baseline CP. The smaller gap between REMI and NB + NMT in the subjective listening test compared to the teacher-forcing NLL evaluation suggests that NB + NMT may be more robust to exposure bias during sequence generation. Another possible explanation is that compound tokens are more effective at capturing the given context, as also demonstrated in the experiments of \cite{hsiao2021compound}.

\begin{table}[]
\centering
\scriptsize
\setlength\tabcolsep{4pt}
\begin{tabular}{lcccc}
\toprule
             & Coherence$\uparrow$     & Richness$\uparrow$       & Consistency$\uparrow$    & Overall$\uparrow$        \\
\midrule
             & \multicolumn{4}{c}{Mean($\pm$margin of error)}    \\
\cmidrule(lr){2-5}
REMI\cite{huang2020pop}         & 3.18 $\pm$ 0.20  & 3.33 $\pm$ 0.18 & 3.33 $\pm$ 0.18 & 3.17 $\pm$ 0.18  \\
CP\cite{hsiao2021compound}   & 2.94 $\pm$ 0.22 & 3.24 $\pm$ 0.18 & 2.97 $\pm$ 0.20 & 3.06 $\pm$ 0.20 \\
\arrayrulecolor{black!40}\cmidrule(lr){1-5}
CP + NMT & 3.22 $\pm$ 0.19 & 3.35 $\pm$ 0.17 & \textbf{3.39} $\pm$ 0.17 & 3.32 $\pm$ 0.17 \\
NB-PF + NMT  & \textbf{3.37} $\pm$ 0.19 & \textbf{3.44} $\pm$ 0.18 & 3.37 $\pm$ 0.19  & \textbf{3.36} $\pm$ 0.20  \\
\arrayrulecolor{black}\bottomrule
\end{tabular}
\caption{Results of subjective listening test, presenting mean values with 95\% confidence intervals.}
\label{table:symbolic_listening_comparison}
\end{table}

% \begin{table}[]
% \caption{Average tokens and standard deviation across datasets, alongside the average number of notes.}
% \centering
% \scriptsize
% \setlength\tabcolsep{3.5pt}
% \begin{tabular}{lcccc}
% \toprule
%                                      & Pop1k7 & POP909 & SOD & Lakh MIDI clean \\
% \midrule
%                                      & \multicolumn{4}{c}{Mean ($\pm$std)} \\
% \cmidrule(lr){2-5}
% REMI & 5,057 $\pm$1,928 & 6,152 $\pm$1,448 & 6,638 $\pm$7,518 & 14,647 $\pm$6,779 \\
% CP   & 2,089 $\pm$771   & 2,709 $\pm$609   & 3,230 $\pm$3,480 & 6,665 $\pm$3,058  \\
% NB   & 1,306 $\pm$547   & 1,659 $\pm$416   & 2,398 $\pm$2,764 & 5,401 $\pm$2,659  \\
% \midrule
% Num. notes                  & 1,303 $\pm$548 & 1,651 $\pm$418 & 2,395 $\pm$3,629 & 5,398 $\pm$2,659 \\
% \bottomrule
% \end{tabular}
% \label{tab:maestro}
% \end{table}

\section{Conclusion}
In summary, this work presents the Nested Music Transformer, an advanced architecture that decodes compound tokens in music generation, applicable to both in the symbolic and audio domain. Our architecture distinguishes itself by addressing the twin challenges of sequence length and feature interdependencies through a nested transformer setup that efficiently manages GPU resources and training processes. The experiments validate the competitiveness of our model over previous methods, achieving on par results in both objective metrics and subjective listening tests while lowering training costs.

\section{Acknowledgements}
This research was supported by the National R\&D Program through the National Research Foundation of Korea (NRF) funded by the Korean Government (MSIT) (RS-2023-00252944, Korean Traditional Gagok Generation Using Deep Learning).

% For bibtex users:
\bibliography{ISMIRtemplate}

@inproceedings{huang2020pop,
  title={Pop Music Transformer: Beat-based Modeling and Generation of Expressive Pop Piano Compositions},
  author={Huang, Yu-Siang and Yang, Yi-Hsuan},
  booktitle={Proceedings of the 28th ACM international conference on multimedia},
  pages={1180--1188},
  year={2020}
}

@inproceedings{zeng2021musicbert,
  author       = {Mingliang Zeng and
                  Xu Tan and
                  Rui Wang and
                  Zeqian Ju and
                  Tao Qin and
                  Tie{-}Yan Liu},
  title        = {MusicBERT: Symbolic Music Understanding with Large-Scale Pre-Training},
  booktitle    = {Findings of the Association for Computational Linguistics: {ACL/IJCNLP}
                  2021, Online Event, August 1-6, 2021},
  series       = {Findings of {ACL}},
  volume       = {{ACL/IJCNLP} 2021},
  pages        = {791--800},
  publisher    = {Association for Computational Linguistics},
  year         = {2021}
}

@inproceedings{dong2023multitrack,
  title={Multitrack music transformer},
  author={Dong, Hao-Wen and Chen, Ke and Dubnov, Shlomo and McAuley, Julian and Berg-Kirkpatrick, Taylor},
  booktitle={ICASSP 2023-2023 IEEE International Conference on Acoustics, Speech and Signal Processing (ICASSP)},
  pages={1--5},
  year={2023},
  organization={IEEE}
}

@inproceedings{hsiao2021compound,
  title={Compound word transformer: Learning to compose full-song music over dynamic directed hypergraphs},
  author={Hsiao, Wen-Yi and Liu, Jen-Yu and Yeh, Yin-Cheng and Yang, Yi-Hsuan},
  booktitle={Proceedings of the AAAI Conference on Artificial Intelligence},
  volume={35},
  number={1},
  pages={178--186},
  year={2021}
}

@inproceedings{wang2020pop909,
  author       = {Ziyu Wang and
                  Ke Chen and
                  Junyan Jiang and
                  Yiyi Zhang and
                  Maoran Xu and
                  Shuqi Dai and
                  Gus Xia},
  title        = {{POP909:} {A} Pop-Song Dataset for Music Arrangement Generation},
  booktitle    = {Proceedings of the 21th International Society for Music Information
                  Retrieval Conference, {ISMIR} 2020, Montreal, Canada, October 11-16,
                  2020},
  pages        = {38--45},
  year         = {2020},
}

@book{raffel2016learning,
  title={Learning-Based Methods for Comparing Sequences, with Applications to Audio-to-MIDI Alignment and Matching},
  author={Raffel, Colin},
  year={2016},
  publisher={Columbia University}
}

@inproceedings{crestel2018database,
  author       = {L{\'{e}}opold Crestel and
                  Philippe Esling and
                  Lena Heng and
                  Stephen McAdams},
  title        = {A Database Linking Piano and Orchestral {MIDI} Scores with Application
                  to Automatic Projective Orchestration},
  booktitle    = {Proceedings of the 18th International Society for Music Information
                  Retrieval Conference, {ISMIR} 2017, Suzhou, China, October 23-27,
                  2017},
  pages        = {592--598},
  year         = {2017}
}

@article{dao2022flashattention,
  title={Flashattention: Fast and memory-efficient exact attention with io-awareness},
  author={Dao, Tri and Fu, Dan and Ermon, Stefano and Rudra, Atri and R{\'e}, Christopher},
  journal={Advances in Neural Information Processing Systems},
  volume={35},
  pages={16344--16359},
  year={2022}
}

@inproceedings{loshchilov2017decoupled,
  author       = {Ilya Loshchilov and
                  Frank Hutter},
  title        = {Decoupled Weight Decay Regularization},
  booktitle    = {7th International Conference on Learning Representations, {ICLR} 2019,
                  New Orleans, LA, USA, May 6-9, 2019},
  publisher    = {OpenReview.net},
  year         = {2019}
}

@article{defossez2022high,
title={High Fidelity Neural Audio Compression},
author={Alexandre D{\'e}fossez and Jade Copet and Gabriel Synnaeve and Yossi Adi},
journal={Transactions on Machine Learning Research},
issn={2835-8856},
year={2023}
}

@article{hawthorne2018enabling,
  title={Enabling factorized piano music modeling and generation with the MAESTRO dataset},
  author={Hawthorne, Curtis and Stasyuk, Andriy and Roberts, Adam and Simon, Ian and Huang, Cheng-Zhi Anna and Dieleman, Sander and Elsen, Erich and Engel, Jesse and Eck, Douglas},
  journal={arXiv preprint arXiv:1810.12247},
  year={2018}
}

@inproceedings{NEURIPS2023_94b472a1,
 author = {Copet, Jade and Kreuk, Felix and Gat, Itai and Remez, Tal and Kant, David and Synnaeve, Gabriel and Adi, Yossi and Defossez, Alexandre},
 booktitle = {Advances in Neural Information Processing Systems},
 editor = {A. Oh and T. Neumann and A. Globerson and K. Saenko and M. Hardt and S. Levine},
 pages = {47704--47720},
 publisher = {Curran Associates, Inc.},
 title = {Simple and Controllable Music Generation},
 volume = {36},
 year = {2023}
}

@inproceedings{gehring2017convolutional,
  title={Convolutional sequence to sequence learning},
  author={Gehring, Jonas and Auli, Michael and Grangier, David and Yarats, Denis and Dauphin, Yann N},
  booktitle={International conference on machine learning},
  pages={1243--1252},
  year={2017},
  organization={PMLR}
}

@inproceedings{gui2024adapting,
  title={Adapting frechet audio distance for generative music evaluation},
  author={Gui, Azalea and Gamper, Hannes and Braun, Sebastian and Emmanouilidou, Dimitra},
  booktitle={ICASSP 2024-2024 IEEE International Conference on Acoustics, Speech and Signal Processing (ICASSP)},
  pages={1331--1335},
  year={2024},
  organization={IEEE}
}

@inproceedings{yanguniaudio,
  title={UniAudio: Towards Universal Audio Generation with Large Language Models},
  author={Yang, Dongchao and Tian, Jinchuan and Tan, Xu and Huang, Rongjie and Liu, Songxiang and Guo, Haohan and Chang, Xuankai and Shi, Jiatong and Bian, Jiang and Zhao, Zhou and others},
  booktitle={Forty-first International Conference on Machine Learning},
  year = {2024}
}

@inproceedings{wu2023large,
  title={Large-scale contrastive language-audio pretraining with feature fusion and keyword-to-caption augmentation},
  author={Wu, Yusong and Chen, Ke and Zhang, Tianyu and Hui, Yuchen and Berg-Kirkpatrick, Taylor and Dubnov, Shlomo},
  booktitle={ICASSP 2023-2023 IEEE International Conference on Acoustics, Speech and Signal Processing (ICASSP)},
  pages={1--5},
  year={2023},
  organization={IEEE}
}

@inproceedings{wang2020pianotree,
  author       = {Ziyu Wang and
                  Yiyi Zhang and
                  Yixiao Zhang and
                  Junyan Jiang and
                  Ruihan Yang and
                  Gus Xia and
                  Junbo Zhao},
  editor       = {Julie Cumming and
                  Jin Ha Lee and
                  Brian McFee and
                  Markus Schedl and
                  Johanna Devaney and
                  Cory McKay and
                  Eva Zangerle and
                  Timothy de Reuse},
  title        = {{PIANOTREE} {VAE:} Structured Representation Learning for Polyphonic
                  Music},
  booktitle    = {Proceedings of the 21th International Society for Music Information
                  Retrieval Conference, {ISMIR} 2020, Montreal, Canada, October 11-16,
                  2020},
  pages        = {368--375},
  year         = {2020},
  url          = {http://archives.ismir.net/ismir2020/paper/000096.pdf},
  bibsource    = {dblp computer science bibliography, https://dblp.org}
}

@inproceedings{huang2019musictransformer,
    title={Music Transformer: Generating Music with Long-Term Structure},
    author={Cheng-Zhi Anna Huang and Ashish Vaswani and Jakob Uszkoreit and Ian Simon and Curtis Hawthorne and Noam Shazeer and Andrew M. Dai and Matthew D. Hoffman and Monica Dinculescu and Douglas Eck},
    booktitle={International Conference on Learning Representations},
    year={2018}
}

@inproceedings{ren2020popmag,
  title={Popmag: Pop music accompaniment generation},
  author={Ren, Yi and He, Jinzheng and Tan, Xu and Qin, Tao and Zhao, Zhou and Liu, Tie-Yan},
  booktitle={Proceedings of the 28th ACM international conference on multimedia},
  pages={1198--1206},
  year={2020}
}

@inproceedings{le2024stack,
  title={Stack-and-delay: a new codebook pattern for music generation},
  author={Le Lan, Gael and Nagaraja, Varun and Chang, Ernie and Kant, David and Ni, Zhaoheng and Shi, Yangyang and Iandola, Forrest and Chandra, Vikas},
  booktitle={ICASSP 2024-2024 IEEE International Conference on Acoustics, Speech and Signal Processing (ICASSP)},
  pages={796--800},
  year={2024},
  organization={IEEE}
}

% For non bibtex users:
%\begin{thebibliography}{citations}
% \bibitem{Author:17}
% E.~Author and B.~Authour, ``The title of the conference paper,'' in {\em Proc.
% of the Int. Society for Music Information Retrieval Conf.}, (Suzhou, China),
% pp.~111--117, 2017.
%
% \bibitem{Someone:10}
% A.~Someone, B.~Someone, and C.~Someone, ``The title of the journal paper,''
%  {\em Journal of New Music Research}, vol.~A, pp.~111--222, September 2010.
%
% \bibitem{Person:20}
% O.~Person, {\em Title of the Book}.
% \newblock Montr\'{e}al, Canada: McGill-Queen's University Press, 2021.
%
% \bibitem{Person:09}
% F.~Person and S.~Person, ``Title of a chapter this book,'' in {\em A Book
% Containing Delightful Chapters} (A.~G. Editor, ed.), pp.~58--102, Tokyo,
% Japan: The Publisher, 2009.
%
%
%\end{thebibliography}

\end{document}